\documentclass[modern]{aastex62}

\usepackage{url}
\usepackage{amsmath}
\usepackage{mathtools}
\usepackage{amssymb}
\usepackage{natbib}
\usepackage{graphicx}
\usepackage{calc}
\usepackage{etoolbox}
\usepackage{xspace}
\usepackage[T1]{fontenc} 
\usepackage{textcomp}
\usepackage{ifxetex}
\ifxetex
\usepackage{fontspec}
\defaultfontfeatures{Extension = .otf}
\fi
\usepackage{fontawesome}
\usepackage{listings}

\newcommand{\starry}{\textsf{starry}\xspace}
\newcommand{\spiceypy}{\textsf{spiceypy}\xspace}
\newcommand{\tf}{\textsf{TensorFlow}\xspace}
\newcommand{\tess}{\emph{TESS}\xspace}
\newcommand{\TESS}{\tess}

\renewcommand{\eqref}[1]{\ref{eq:#1}}

\definecolor{linkcolor}{rgb}{0.1216,0.4667,0.7059}
\newcommand{\codeicon}{{\color{linkcolor}\faFileCodeO}}

\newcommand{\codelink}[1]{\href{https://github.com/rodluger/earthshine/blob/682386970078e42a09d7a45f1e242dc0716820c1/notebooks/#1.ipynb}{\codeicon}\,\,}

\newtagform{eqtag}[]{(}{)}
\newcommand{\currentlabel}{None}

\definecolor{codegreen}{rgb}{0,0.6,0}
\definecolor{codegray}{rgb}{0.5,0.5,0.5}
\definecolor{codepurple}{rgb}{0.58,0,0.82}
\definecolor{backcolour}{rgb}{0.95,0.95,0.95}
\lstdefinestyle{mystyle}{
    backgroundcolor=\color{backcolour},
    commentstyle=\color{codegreen},
    keywordstyle=\color{magenta},
    numberstyle=\tiny\color{codegray},
    stringstyle=\color{codepurple},
    basicstyle=\small\ttfamily,
    breakatwhitespace=false,
    breaklines=true,
    captionpos=b,
    keepspaces=true,
    numbers=left,
    numbersep=5pt,
    showspaces=false,
    showstringspaces=false,
    showtabs=false,
    tabsize=2,
    aboveskip=1em,
    belowskip=1em,
    keywords=[2]{map},
    keywordstyle=[2]{\color{black!80!black}},
    upquote=true
}
\renewcommand\quad{\hskip\fontdimen3\font}

\usepackage{xcolor}

\begin{document}


\title{\TESS Photometric Mapping of a Terrestrial Planet in the Habitable Zone: 
       Detection of Clouds, Oceans, and Continents}

\author[0000-0002-0296-3826]{Rodrigo Luger}
\email{rluger@flatironinstitute.org}
\affil{Center~for~Computational~Astrophysics, Flatiron~Institute, New~York, NY}
\author[0000-0002-9328-5652]{Megan Bedell}
\affil{Center~for~Computational~Astrophysics, Flatiron~Institute, New~York, NY}
\author{Roland Vanderspek}
\affil{Kavli Institute for Astrophysics and Space Research, Massachusetts 
       Institute of Technology, Cambridge, MA}
\author{Christopher J. Burke}
\affil{Kavli Institute for Astrophysics and Space Research, Massachusetts 
       Institute of Technology, Cambridge, MA}

\begin{abstract}
To date, a handful of exoplanets have been photometrically mapped using phase-modulated 
reflection or emission from their surfaces, but the small amplitudes 
of such signals have limited previous maps almost exclusively to coarse dipolar
features on hot giant planets. 
In this work, we uncover a signal using recently released data from the Transiting 
Exoplanet Survey Satellite (\TESS), which we show corresponds to time-variable reflection 
from a terrestrial 
planet with a rotation period of $0.9972696$ days. 
Using a spherical harmonic-based reflection model developed as an extension of 
the \starry package, we are able to reconstruct the surface features of this rocky 
world. We recover a time-variable albedo map of the planet including persistent 
regions which we interpret as oceans and cloud banks indicative of continental features. 
We argue that this planet represents the most promising detection of a habitable 
world to date, although the potential intelligence of any life on it is yet to 
be determined.
\href{https://github.com/rodluger/earthshine}{\color{linkcolor}\faGithub}
\end{abstract}


\section{Introduction}
\label{sec:intro}

Of the many challenges and opportunities facing the field of exoplanets in the future, 
the prospect of mapping the surface of a potentially habitable planet is one of 
the most exciting -- and one of the most difficult. 
Present-day missions including the Transiting Exoplanet Survey Satellite 
\citep[\TESS; ][]{Ricker2015} aim to discover small terrestrial planets around 
bright, nearby stars. 
Once these targets are found, more detailed observational characterization 
must follow to evaluate their suitability for harboring life. 
The discovery of surface features such as continents and oceans would be a 
critical step toward understanding the habitability of an alien world.

The most straightforward method of observing features on a terrestrial exoplanet 
would be to directly image the planet at sufficient resolution. 
However, such an observation would require an effective telescope aperture many orders 
of magnitude larger and equipped with far more advanced coronagraphic optics
than what we have available today.
Fortunately, surface mapping can also be done indirectly by observing rotational 
phase-dependent variations of the planetary signal in precise photometric time series. 
This is a complex data analysis problem: strong degeneracies are involved when 
transforming a 1-dimensional time series into a 2-dimensional surface map
\citep{CowanFuentesHaggard2013}. 
Interpretation is also challenging in the face of many potential explanations 
for a poorly resolved surface feature. Nevertheless, this technique has
already proven successful in the mapping of large dipolar features on the surfaces
of hot giant planets, most notably the hot Jupiter HD189733b
\citep{Knutson2007,Majeau2012,deWit2012}. More recently, this technique has also
been applied to the potentially terrestrial planet 55 Cancri e, suggesting
a large longitudinal offset in its peak surface emission due to either
lava flows on the surface or strong atmospheric recirculation
\citep{Demory2016,Demory2016b,Hammond2017}.

However, obtaining a map of a temperate terrestrial exoplanet is still
beyond the capabilities of current facilities, as both the thermal and 
reflected light signals from these planets
are orders of magnitude below current measurement capabilities. Nevertheless,
much work has been done on the theoretical front on mapping terrestrial
planets in the habitable zone
\citep[e.g.,][]{Kawahara2010,Fujii2012,Berdyugina2017,Haggard2018,Lustig-Yaeger2018};
for a review, see \cite{Cowan2018}.

In this work, we show how we can harness photometric data collected by the \TESS mission
to construct a map of Sol d, a rocky planet with a radius of $6.4\times 10^6$~m
orbiting near the inner edge of the habitable zone of its star, a nearby G2 main sequence
star \citep{Sagan1993}. Although the \TESS mission promised to deliver a handful of terrestrial planets in 
the habitable zone for follow-up with other missions, it came as a great surprise
that data from the spacecraft could be used not only to detect Sol d but to
characterize it in detail. This is due primarily to the strong nature of the reflected
light signal originating from the planet, which 
because of complex internal optics illuminates the entirety of the \TESS detector 
during most of Sector 1 and part of Sector 2.

Previous attempts to map Sol d from the exoplanet community have suggested 
the presence of localized surface features, but their resolving power has been 
limited by the cadence and/or duration of observations \citep{Cowan2009,Jiang2018}. 
\TESS offers precise 2-minute cadence data spanning a wide range of illumination 
phases, enabling a level of spatial and temporal resolving power that is 
unprecedented in the history of exoplanet phase mapping. 
Moreover, recent advances on the modeling side 
have made the reconstruction of detailed maps more feasible than ever. 
\cite{Luger2019} developed an analytic algorithm for generating---and
inverting---light curves of stars and planets whose surface emission is
described in terms of spherical harmonics, dubbed \starry. The analyticity of this
method makes the computation of the model both fast and precise.
In turn, \cite{Haggard2018} derived analytic expressions for the
analogous case in reflected light, packaged into the 
\textsf{EARL} code. In this work, we derive similar expressions within
the \starry framework to perform fast, analytical inversion of the light curve
of Sol d and thus construct a map of the planet in reflected light.

The paper is organized as follows: we give an overview of the \TESS data used in \S\ref{sec:data},
and in \S\ref{sec:methods} we describe the methods employed to infer a map 
from these data, including the adaptation of the \starry algorithm to reflection 
mapping, the modeling of spacecraft-related systematics, and the likelihood and 
priors used. 
We present results in \S\ref{sec:results} and make a comparison of these 
findings to other state-of-the-art maps of Sol d in 
\S\ref{sec:discussion}. 
Finally, we conclude with a summary of our major findings in \S\ref{sec:conclusions}.
The code used to generate all the figures in this paper is open source
and available on GitHub%
\footnote{\url{https://github.com/rodluger/earthshine}}. Clickable icons
\href{https://github.com/rodluger/earthshine}{\color{linkcolor}\faFileCodeO}
next to each figure caption 
link to the source code used to generate them.

\section{Data}
\label{sec:data}

The \TESS pipeline produces processed light curves for all short-cadence (2 minute) targets. 
Part of this processing is the determination of the localized background flux 
within each postage stamp, which is provided as an ancillary time series. 
As with all \TESS time series data products, the time is given in units of \TESS Julian 
Date (TJD), defined as $\mathrm{BJD} - 2457000$, and the data are delivered in 27-day 
parcels corresponding to the varying pointing sectors of \TESS. 
We use these background flux measurements from Sectors 1 and 2 as our primary 
data source in this work. 

For each of the two sectors considered, we downloaded a random subsample of 
1,000 short-cadence light curve files corresponding to postage stamps located 
across the full \TESS field of view and extracted the background flux time 
series (\textsf{SAP\_BKG}) and corresponding uncertainties
(\textsf{SAP\_BKG\_ERR}) from each. 
We then interpolated all light curves in a given sector onto the same time
grid, masked all cadences with nonzero \textsf{QUALITY} flags, and masked
cadences during which Sol d was below the edge of the sunshade.

\begin{figure}[t!]
    \begin{centering}
    \includegraphics[width=\linewidth]{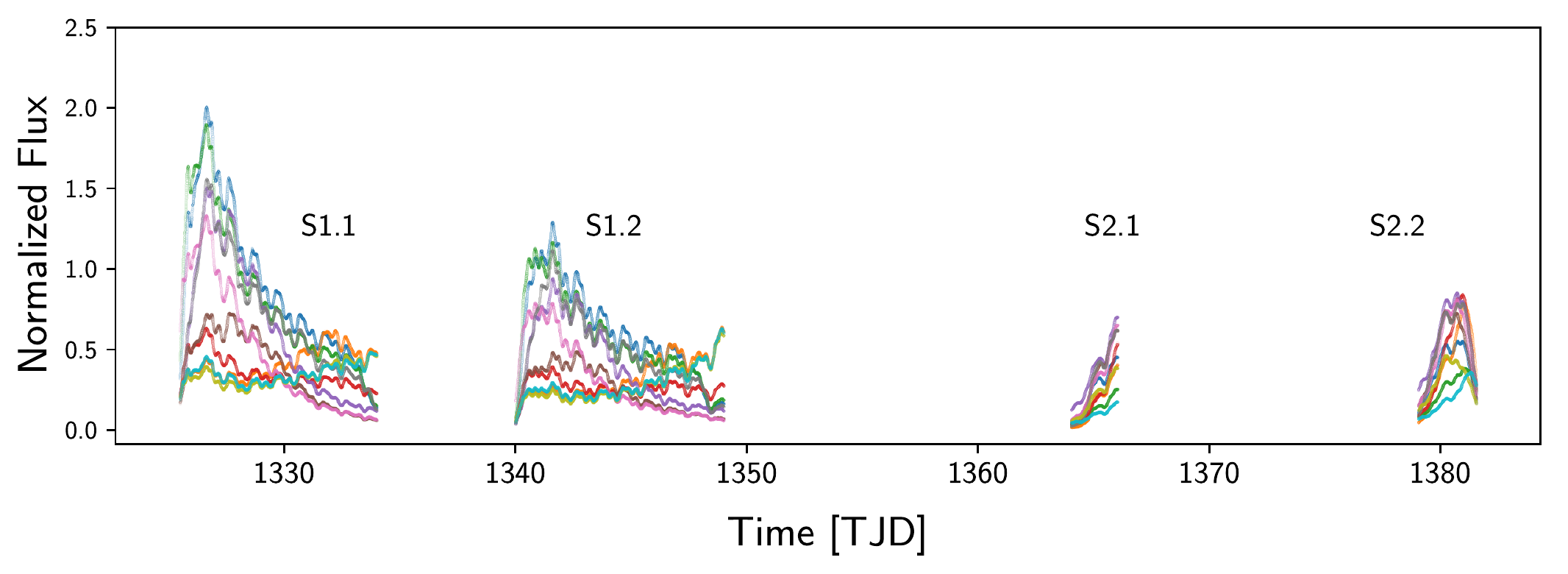}
    \caption{\label{fig:data}
             Normalized background flux in each of ten postage stamps
             for each of Sectors 1 and 2. Cadences where Sol d was
             below or very close to the edge of the sunshade were masked.
             While the planet is in view for the majority of both orbits
             in Sector 1, it is only visible in the data at the very end
             of each orbit in Sector 2.
             \codelink{earthshine_S1_S2}
             }
    \end{centering}
\end{figure}

Not all postage stamps are equally informative, since the desired signal from 
Sol d is inhomogeneously spread across the \TESS detectors and manifests itself
differently in different regions (see \S\ref{sec:systematics}). 
We down-selected our pool of 1,000 targets to 75 maximally useful background 
light curves from Sector 1 and 20 from Sector 2. 
This down-selection was done with several criteria in mind: each light curve 
should contain a significant signal from Sol d; they should be representative 
of the general background flux modulation seen across the \TESS cameras, rather 
than containing any truly localized signals from astrophysical background 
sources; and the sample as a whole should contain the full range of typical 
behaviors of background across the entire \TESS field of view. 
To accomplish this, we applied a categorization and outlier rejection scheme to the ensemble of light curves.
During this selection step only, we normalized the light curves to the same min-max range.
We then computed the median
flux at each cadence along with an estimate of the variance based on the
median absolute deviation (MAD) across all light curves.
For each light curve, we calculated a score $\chi^2$ equal to the sum of the squares of the
deviation from the median at each of the $N$ cadences, normalized by the variance.
We removed all light curves with $\chi^2 < N$ from the pool, placing
them in a separate group. We repeated these steps several times,
classifying light curves into four groups for Sector 1 and two groups for Sector 2. 
We found that this
procedure effectively grouped together light curves with similar features,
increasing in variation within each group from the first group (most homogeneous) 
to the last group (most heterogeneous). In practice, we found light curves
in all but the last group to be dominated by 
the signal of Sol d, while light
curves in the last group were dominated by signals other than the
periodic modulation of the planet. To obtain our final set of light curves, 
we sorted the light curves in each group by the amplitude
of the signal and kept the first 25 in each of the first three groups in Sector 1
and the first 20 in the first group in Sector 2 for a total
of 75 and 20 light curves, respectively. We then computed an estimate of the
non-Sol d background flux from the cadences during which the planet was below
the edge of the sunshade and removed this baseline from the flux.
Finally, we median-normalized all light curves (as opposed to the min-max
normalization used for the classification).
Our final dataset consisted of 912,605 data points in total. 
A representative sample of these signals is shown in Figure~\ref{fig:data}.

\pagebreak

\section{Methods}%
\label{sec:methods}%
\subsection{Orbital Modeling}%
\label{sec:orbit}%
We downloaded%
\footnote{\url{https://archive.stsci.edu/missions/tess/models/}}
the Solar System and \TESS ephemeris data for
Sectors 1 and 2 and used \spiceypy \citep{Acton1996, Acton2017, Annex2017}
to convert the TJD timestamps to 
ephemeris time (ET). We then used the \textsf{spkezr} utility of \spiceypy to compute
the relative positions of Sol, Sol d, and \TESS for every cadence
in our dataset in the J2000 equatorial reference frame. In this right-handed
coordinate frame, Sol d is centered at the origin, the $y$ axis points along the 
planet's spin axis, and the $x$ axis points toward the vernal equinox. We
do not apply any light travel correction, as the expected delay is on the order
of a few seconds.

At any given cadence, we rotate Sol d about its spin axis to the correct phase,
assuming a planetary rotation period of $0.9972696$ days and an obliquity
of $23.437^\circ$. We determine the initial rotation
phase of Sol d by computing the Sol d Rotation Angle (ERA), given by \citep{Urban2013}
\begin{align}
\mathrm{ERA} = 360^\circ(0.7790572732640 + 1.00273781191135448 \, t_U) \, \mathrm{mod} \, 360^\circ
\end{align}
where $t_U = t - 2451545.0$. At $t = t_0 = 2458325.5$, the first timestamp in our dataset,
we find $\mathrm{ERA} = 303.4^\circ$, meaning the vernal equinox will be aligned with
the prime meridian $360^\circ - 303.4^\circ = 56.6^\circ = 3.77$ hours past $t_0$.
Finally, we rotate the frame to align
\TESS with the $+z$ axis, with the spin vector of Sol d along the $y-z$ plane.
We verified our rotations by comparing our results on several days to the
ephemerides provided by the JPL HORIZONS Web interface%
\footnote{See \url{https://ssd.jpl.nasa.gov/horizons.cgi} and 
\href{https://github.com/rodluger/earthshine/blob/master/notebooks/SanityCheck.ipynb}{this
Jupyter notebook.}}.

\begin{figure}[t!]
    \begin{centering}
    \includegraphics[width=\linewidth]{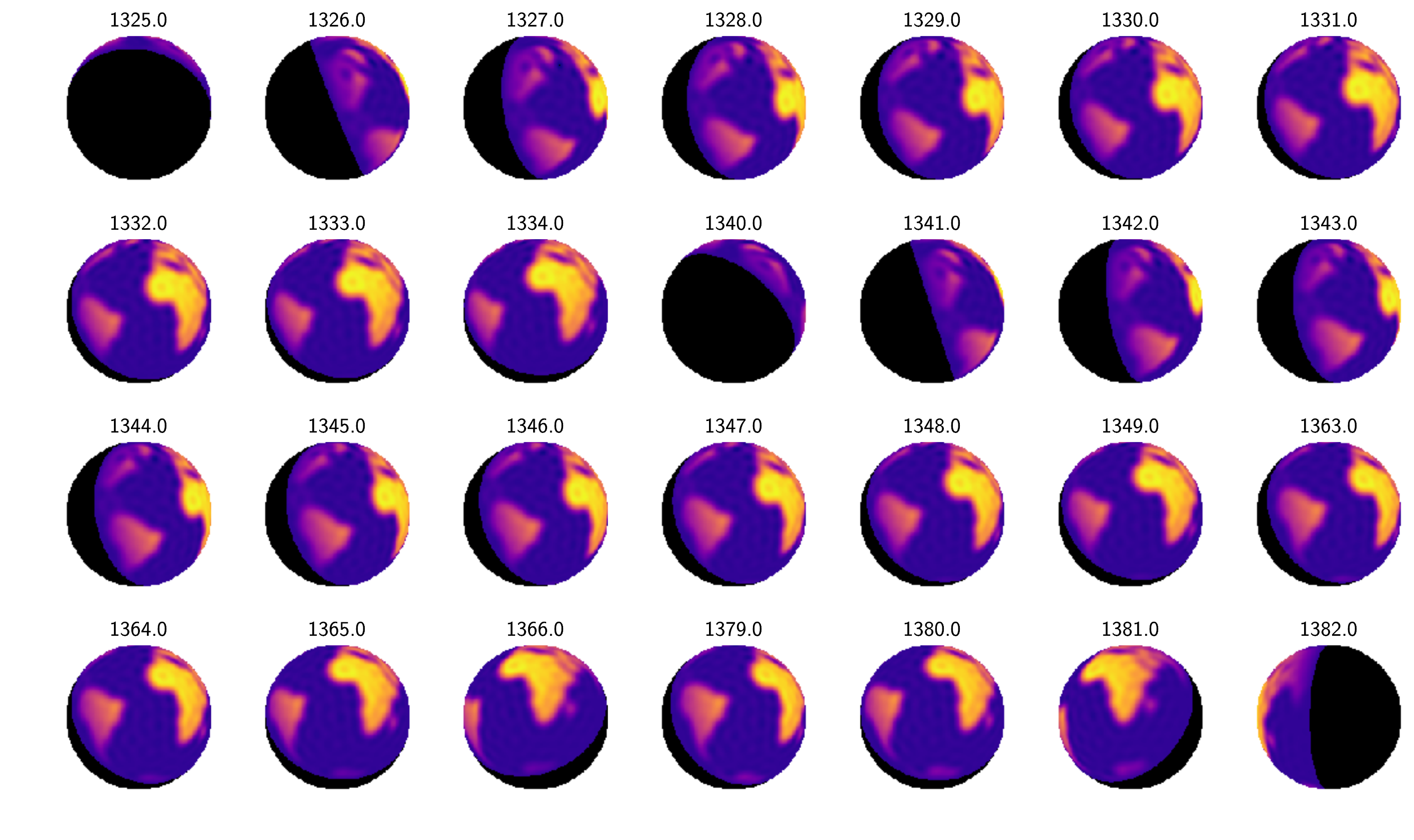}
    \caption{\label{fig:phases}
             Synthetic images showing \TESS's view of Sol d on every date in 
             Sectors 1 and 2 that was included in the regression. Each image 
             is labeled with the corresponding timestamp in TJD. 
             The dayside features shown here correspond to a hypothetical 
             model of Sol d, not to the map inferred in this study.
             \codelink{EarthView}
             }
    \end{centering}
\end{figure}

Figure~\ref{fig:phases} shows the view \TESS has of Sol d on each of the dates
in our dataset; these images were produced by applying the sequences of rotations
described above. The planet is illuminated by Sol assuming it is a point source,
and the night side is shaded black. The map of Sol d shown in this figure is 
based on current models of the actual distribution of continents on the planet.
Note that although the images are computed an integer number of rotation periods
apart, the planet appears to slowly rotate over the course of the observation.
This is due to the orbit of \TESS, which changes its vantage point relative
to Sol d over the course of its 13 day orbit.

\subsection{Systematics Modeling}
\label{sec:systematics}

While the astrophysical signal that we wish to analyze is present in all 
extracted light curves, its strength is modulated by time-variable systematics 
that are correlated but not identical across all postage stamps, since the
illumination of the CCD by Sol d is strongly inhomogeneous. A physical
model of the reflections and scattering that gives rise to the
signal of Sol d on the \TESS detector is beyond the scope of this work. Instead,
we treat this as a de-trending problem, where the signal we wish to fit is
shared by all postage stamps, but multiplied by a systematic signal that is
different for each target. This problem is precisely what the technique of
pixel-level decorrelation \citep[PLD;][]{Deming2015, Luger2016, Luger2018a}
seeks to address. In PLD, one computes the sum over all $N$ light curves at each
point in time, and uses this quantity to normalize each of the individual
light curves. Since (by assumption) each light curve is the product of the astrophysical
signal and the systematics signal, this procedure divides out the astrophysical
signal, producing a basis of $N$ vectors that contain \emph{only} functions of the
systematics. This now serves as a ``clean'' basis one can then use to fit the 
systematics component of the data, with minimal risk of fitting out astrophysical signals. 

We use PLD to determine a systematics basis set for each \TESS sector, and
from this basis we construct a separate design matrix $\mathbf{B}$ for each sector. To
increase the flexibility of the systematics model, we append to $\mathbf{B}$ a basis of
fifth order orthogonal polynomials in time for each of the two orbits in each sector.
The systematics model $\mathbf{p}$ for the $n^\mathrm{th}$ signal is thus
\begin{align}
\mathbf{p}_n = \mathbf{B} \mathbf{w}_n
\end{align}
where $\mathbf{w}_n$ are the weights of each regressor.
Since we solve for different weights for each of the 75 light curves in Sector 1
and 20 light curves in Sector 2, our systematics model has
$75 \times (75 + 2 \times 5) + 20 \times (20 + 2 \times 5) = 6,975$ free parameters.
The total number of data points is $912,605$, so overfitting is not particularly
concerning; nevertheless, we impose an L2 prior on the weights to minimize this risk
(see \S\ref{sec:model}).

\subsection{Reflected Light Mapping with \starry}
\label{sec:starry}

We model the astrophysical component of the signal---the rotational modulation
of Sol d in reflected light---using the \starry code package \citep{Luger2019}.
\starry computes phase curves and occultation light curves for bodies whose
surfaces can be expressed as a sum of spherical harmonics. 
The algorithm works by first transforming
from spherical harmonic coefficients to coefficients in the polynomial basis $\tilde{p}$, whose
terms are of the form $x^i y^j z^k$ for (non-negative) integer $i$ and $j$ 
and $k = 0$ or $1$. \citet{Luger2019} showed how to compute the visible
flux by integrating each of the terms in $\tilde{p}$ across the visible
surface of the projected disk. For unocculted spheres observed in emitted
light, the area of integration is the full disk, which makes the integration
trivial: the flux contribution from each term in $\tilde{p}$ is a constant
easily computed via recursion relations involving factorials. The algorithm is
thus extremely fast and very precise.

However, in the present work, we are interested in \emph{reflected} light, so we must
account for the illumination by the star. This illumination varies smoothly
across the dayside of the planet, but its derivative discontinuously changes at the 
terminator, beyond which the illumination is zero everywhere. This requires
two modifications to the \starry algorithm: we must (1) weight all integrands
by the illumination function on the dayside, and (2) modify the integration
limits to truncate the integrals at the day/night terminator.

\begin{figure}[t!]
    \begin{centering}
    \includegraphics[width=0.4\linewidth]{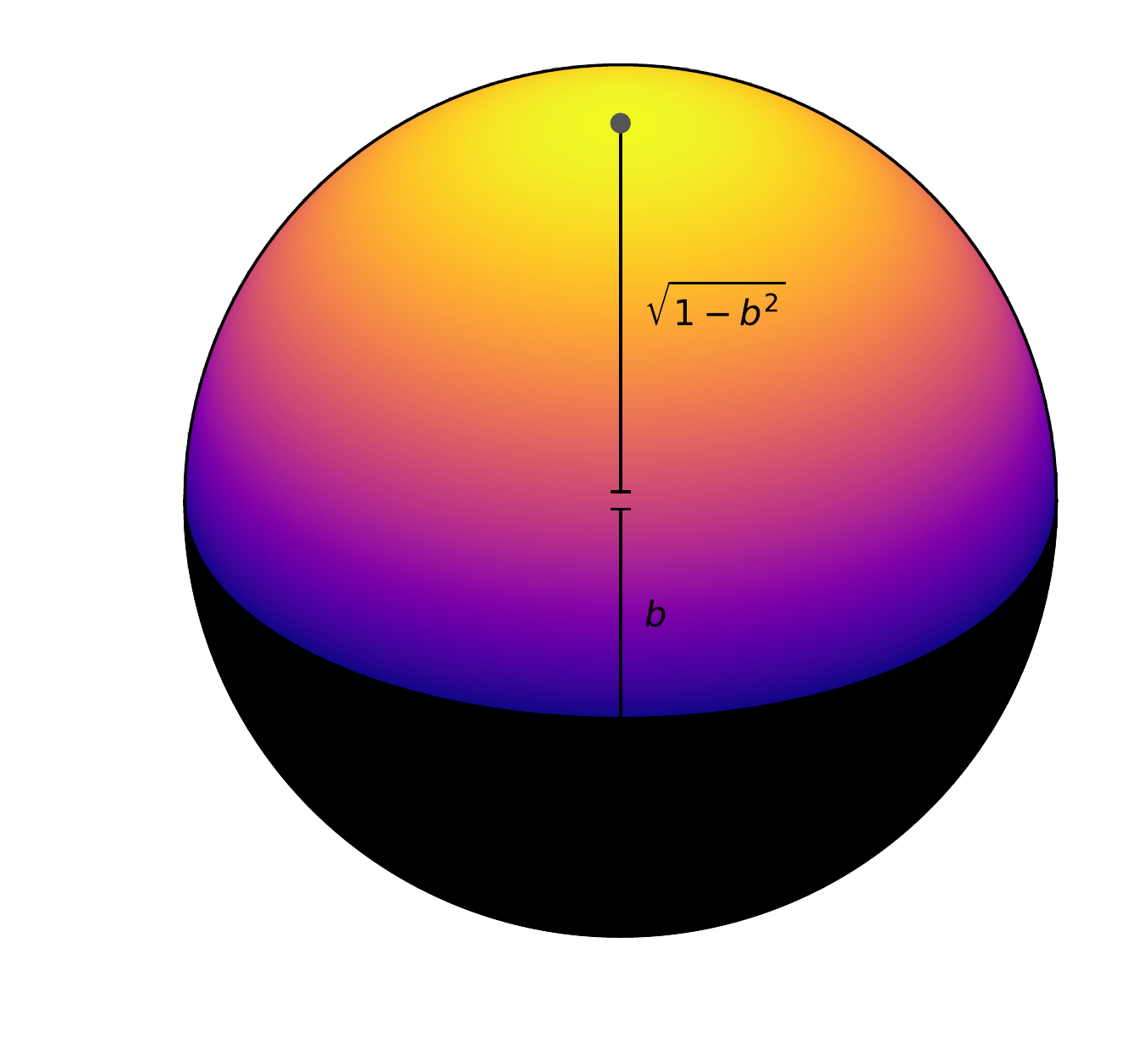}
    \caption{\label{fig:geometry}
             Geometry of the \starry model in reflected light. In this frame, 
             the $y$ axis points up, 
             the $x$ axis points to the right, and the $z$ axis points out of the page.
             The planet has unit radius and sits at the origin, while
             the illumination source is along the $y-z$ plane, with the sub-stellar point is
             marked by a dot. The semi-major axis of the terminator is unity, and
             the semi-minor axis of the terminator is denoted $b$. The night side
             is colored black.
             \codelink{Geometry}
             }
    \end{centering}
\end{figure}

Let us consider a right-handed coordinate frame in which the planet has unit 
radius and is located at the origin, the illumination source is along the $y-z$
plane, and the sub-stellar point is on the $+y$ axis at $x = 0$
(see Figure~\ref{fig:geometry}). In this frame, the terminator is a segment of
an ellipse, with semi-major axis $a=1$ aligned with the $x$ axis and
semi-minor axis $b$. It is straightforward to show that the sub-stellar point
is located at $y=\sqrt{1 - b^2}$ and the illumination is given by
\begin{align}
    I(b; x, y) &= 
    \begin{dcases}
        \sqrt{1 - b^2}y - bz & 
            \quad\quad\quad\quad\quad\quad\quad\quad\quad\quad 
            y \geq -b \sqrt{1 - x^2}
        \\
        0 & 
            \quad\quad\quad\quad\quad\quad\quad\quad\quad\quad 
            \mathrm{otherwise}
    \end{dcases}
    \quad ,
    \label{eq:illum}
\end{align}
where $z = \sqrt{1 - x^2 - y^2}$.
Since the illumination function is just a polynomial in $x$, $y$, and $z$,
weighting the terms in $\tilde{p}$ by this function keeps all terms within the
polynomial basis.
Provided we always rotate the problem such that the
sub-stellar point lies along the $+y$ axis as above, the limits of integration
are $-1 \leq x \leq 1$ and $b\sqrt{1 - x^2} \leq y \leq \sqrt{1 - x^2}$.
Computing the total reflected flux for any surface map is therefore a matter
of solving integrals of the form
\begin{align}
    J_{ijk}(b) &= \int_{-1}^{1} \int_{b\sqrt{1 - x^2}}^{\sqrt{1 - x^2}} x^i y^j z^k \mathrm{d} y \ \mathrm{d} x
    \nonumber\\
    &=
    \frac{
        \Gamma\left(\frac{i + 1}{2}\right) \Gamma\left(\frac{j + 1}{2}\right)
    }{
        2 \Gamma\left(\frac{i + j + k + 4}{2}\right)
    }
    \begin{dcases}
        0
        &
        \quad\quad\quad\quad\quad\quad\quad\quad\quad\quad 
        i \, \mathrm{odd}
        \\
            1 - b^{j + 1}
        & 
        \quad\quad\quad\quad\quad\quad\quad\quad\quad\quad 
        k = 0
        \\
        \frac{\sqrt{\pi}}{2} - b^{j + 1}\Gamma\left(\frac{j + 4}{2}\right)
        {_2\tilde{F}_1} \left( -\frac{1}{2}, \frac{j + 1}{2}; \frac{j + 3}{2}; b^2\right)
        & 
        \quad\quad\quad\quad\quad\quad\quad\quad\quad\quad 
        k = 1
    \end{dcases}
    \quad ,
    \label{eq:integral}
\end{align}
where ${_2\tilde{F}_1}$ is the regularized hypergeometric function. For all values
of $i$, $j$, and $k$ in $\tilde{p}$, ${_2\tilde{F}_1}$ reduces to simple
trigonometric functions involving $b$. Moreover, it is straightforward to
derive recursion relations for $J_{ijk}(b)$, making its evaluation extremely fast.
We implemented the algorithm for computing $J_{ijk}(b)$ in the development version
of \starry
\footnote{\url{https://github.com/rodluger/starry/tree/linear}};
the details will be discussed in more detail in upcoming work.

\begin{figure}[t!]
    \begin{centering}
    \includegraphics[width=\linewidth]{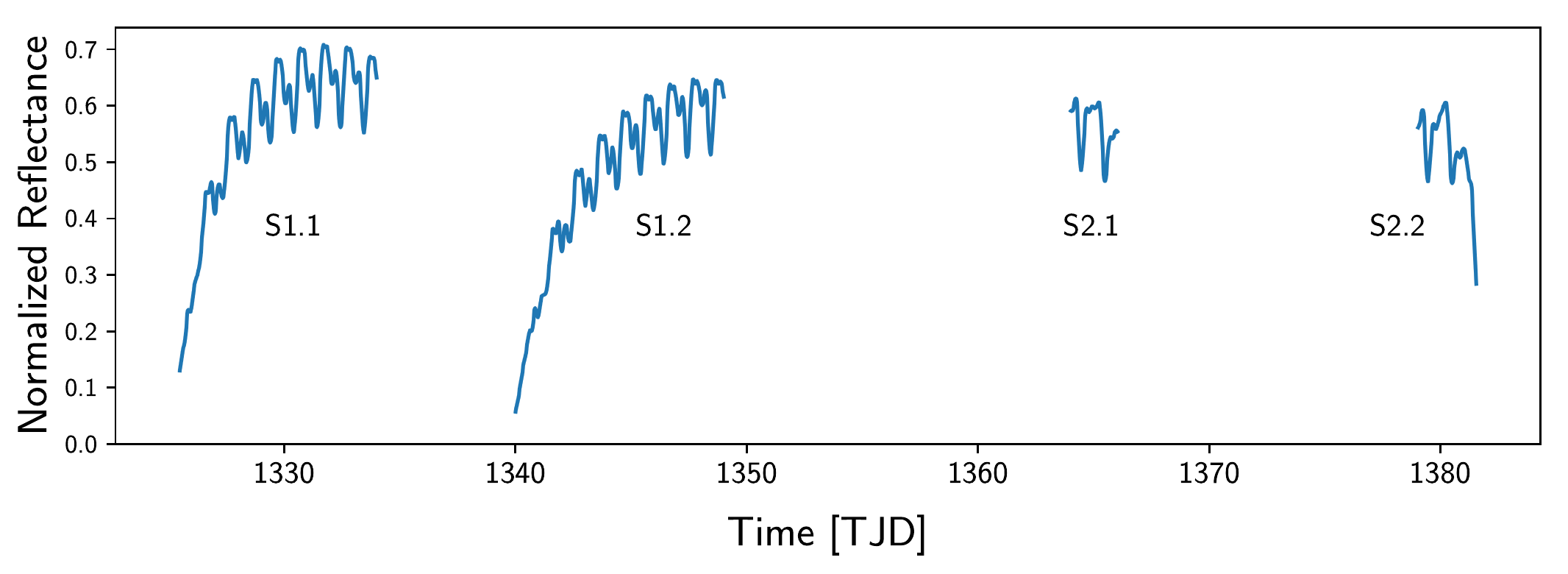}
    \caption{\label{fig:starry_model}
             Normalized surface reflectance versus time for the maximum
             likelihood \starry model in each of the four orbits of \TESS
             in Sectors 1 and 2.
             \codelink{earthshine_S1_S2}
             }
    \end{centering}
\end{figure}

As with the systematics model (\S\ref{sec:systematics}), the model
for the astrophysical signal is \emph{linear}. The signal $\mathbf{s}$ 
is a linear operation on the vector of spherical harmonic coefficients 
$\mathbf{y}$
that describes the body's surface:
\begin{align}
    \mathbf{s} = \mathbf{A} \mathbf{y}
\end{align}
where $\mathbf{A}$ is the design matrix and $\mathbf{y}$ is the vector of spherical harmonic coefficients. 
Note that in \cite{Luger2019}, $\mathbf{y}$ describes the emissivity of the surface,
but here we will use it to represent the \emph{albedo} of the planet.

We chose a maximum spherical harmonic degree $l_{max} = 10$ for our fits.
Increasing $l_{max}$ above this value 
leads to a sharp rise in the amount of ``ringing'' without substantially
increasing the quality of the fit. Note that $l = 10$ corresponds to
a maximum surface resolution on the order of $180^\circ/10 \approx 18^\circ$.

Finally, for reference, in Figure~\ref{fig:starry_model} we plot the
\starry model for our maximum likelihood fit (\S\ref{sec:results}). The
sharp rise at the beginning of each orbit in Sector 1 is due to the changing
vantage point of \TESS as Sol d transitions from a thin crescent to full phase.

\subsection{Full Model and Likelihood Function}
\label{sec:model}

Our full model for each light curve is simply the product of the 
systematics model $\mathbf{p}_n = \mathbf{B}\mathbf{w}_n$ (different for each target) and
the \starry model $\mathbf{s} = \mathbf{A}\mathbf{y}$ (shared by all targets). Since we happen
to know the exact distance $\mathbf{r}$ between \TESS and Sol d at all times
(\S\ref{sec:orbit}), we also include the inverse square of $\mathbf{r}$ 
as a multiplicative term to scale the luminosity 
of Sol d to the flux observed by \TESS. The model for the flux time series
in the $n^\mathrm{th}$ postage stamp is therefore
\begin{align}
    \label{eq:model}
    \mathbf{m}_n = (\mathbf{r}^{-2}) \circ (\mathbf{B} \mathbf{w}_n) \circ (\mathbf{A} \mathbf{y})
\end{align}
where $\circ$ denotes the element-wise product of two vectors.

We solve Equation~(\ref{eq:model}) for the weights $\mathbf{w}_n$ and $\mathbf{y}$
by maximizing the negative log likelihood function in two separate steps. In the
first step, we take advantage of the linearity of the problem to perform a fast,
semi-analytical optimization to obtain a starting guess for the weights. In the
second step, we apply additional constraints to prevent overfitting and ensure
an albedo in the range $[0, 1]$ everywhere on the surface.

In the first step, we initialize the \starry model to unity and iteratively
solve for $\mathbf{w}_n$ and $\mathbf{y}$ by solving the L2-regularized
least squares problem \citep[see, e.g., \S2.1 in][]{Luger2018a}.
We assume zero-mean Gaussian priors for the weights:
\begin{equation}
    \label{eq:wprior}
    \begin{aligned}
        \mathbf{w}_n \sim \mathcal{N}(0, \sigma_w^2)
    \end{aligned}
    \qquad\qquad\qquad\qquad
    \begin{aligned}
        \mathbf{y} \sim \mathcal{N}(0, \sigma_y(l)^2)
    \end{aligned}\, ,
\end{equation}
with $\sigma_w = 0.05$ and 
\begin{align}
    \sigma_y(l) &=
    \begin{dcases}
        1.75\times 10^{-5} \, l^\frac{3}{2} & 
            \quad\quad\quad\quad\quad\quad\quad\quad\quad\quad 
            l < 2
        \\
        1.40\times 10^{-4} \, l^{-\frac{3}{2}} & 
            \quad\quad\quad\quad\quad\quad\quad\quad\quad\quad 
            l \geq 2
    \end{dcases}
    \, ,
\end{align}
where $l$ is the spherical harmonic degree. The latter prior was chosen
by trial-and-error and was adjusted to keep the albedo positive
everywhere on the surface. Note, importantly, that we do
not fit for the coefficient of the constant $Y_{0,0}$ spherical harmonic,
but instead fix it to unity. Since the solutions obtained in this step are merely
used as a starting point for the second step (see below), the choice of prior 
at this stage does not significantly impact on our results.

In general, we find that the iterative scheme converges within about 50 
iterations, taking about two minutes on a laptop computer. However, since it 
is strongly regularized, the model significantly
underfits the data. In the second step, we remove our L2 prior on the spherical
harmonic coefficients and instead regularize the actual surface albedo, which
we compute on a uniform spherical grid with 50,000 points. We enforce a 
uniform prior in the range $[0, 1]$, with a small amount of Gaussian smoothing on either 
side. We additionally impose a constraint on the power spectrum of the
inferred map, requiring that it be drawn from a distribution whose mean
is a decaying power law in the spherical harmonic degree $l$ in order
to suppress ringing and spurious high order features.
Our full negative log likelihood function in this step is therefore
\begin{align}
    \label{eq:nll}
    -\log\mathcal{L} \, = \, 
        &\frac{1}{2}(\mathbf{f} - \mathbf{m})^\top \boldsymbol{\Sigma}^{-1} (\mathbf{f} - \mathbf{m}) \, + \nonumber \\
        &\frac{1}{2}\mathbf{w}^\top \boldsymbol{\Lambda}_w^{-1} \mathbf{w} \, + \nonumber \\
        &\frac{1}{2}(\mathbf{a_-} + 1 - \mathbf{a_+})^\top \boldsymbol{\Lambda}_a^{-1} (\mathbf{a_-} + 1 - \mathbf{a_+}) \, + \nonumber \\
        &\frac{1}{2}(\boldsymbol{\rho} - \boldsymbol{l}^{-\gamma})^\top \boldsymbol{\Lambda}_\rho^{-1} (\boldsymbol{\rho} - \boldsymbol{l}^{-\gamma}) \, + \nonumber \\
        &\frac{1}{2} \frac{(\gamma - \gamma_0)^2}{\sigma_\gamma^2}
        \quad,
\end{align}
where $\mathbf{f}$ is the measured flux across all targets,
$\mathbf{m}$ is the full model (Equation~\ref{eq:model}),
$\boldsymbol{\Sigma}$ is the (diagonal) data covariance, 
$\mathbf{w}$ are the systematics model weights for all targets,
$\boldsymbol{\Lambda}_w$ is the prior covariance of the weights
(Equation~\ref{eq:wprior}), $\mathbf{a_-}$ and $\mathbf{a_+}$ are
vectors containing the values of the albedo that are below zero and above one, 
respectively, $\boldsymbol{\Lambda}_a$ is the covariance of the Gaussian
used to smooth the edges of the top-hat prior on $\mathbf{a}$,
$\boldsymbol{\rho}$ is the power spectrum of the inferred spherical harmonic map
(equal to the sum of the squares of the coefficients at each degree $\boldsymbol{l}$),
$\gamma$ is the power law index of the power spectrum prior, $\boldsymbol{\Lambda}_\rho$ is the
covariance of that prior, $\gamma_0 = 1.5$ is the mean of the prior on $\gamma$
and $\sigma_\gamma^2 = 10^{-2}$ is the corresponding variance.
For simplicity, all our prior covariances are diagonal and
homoscedastic with variances equal to $2.5\times 10^{-3}$ for $\mathbf{w}$,
$10^{-8}$ for $\mathbf{a}$, and $10^{-1}$ for $\boldsymbol{\rho}$.

Unfortunately, the constraints outlined above break the linearity 
of the problem, and we must turn to a non-linear optimizer. We use the 
\textsf{AdamOptimizer} method in the \tf package \citep{Abadi2015}
to maximize Equation~(\ref{eq:nll}). Starting with the values of the
weights obtained in the first step, we run the \textsf{AdamOptimizer}
for 1000 iterations, well past convergence. Since \tf automatically
computes and propagates gradient information, this step is fast, 
finishing in under one minute on a laptop.

\section{Results}
\label{sec:results}

\begin{figure}[t!]
    \begin{centering}
    \includegraphics[width=\linewidth]{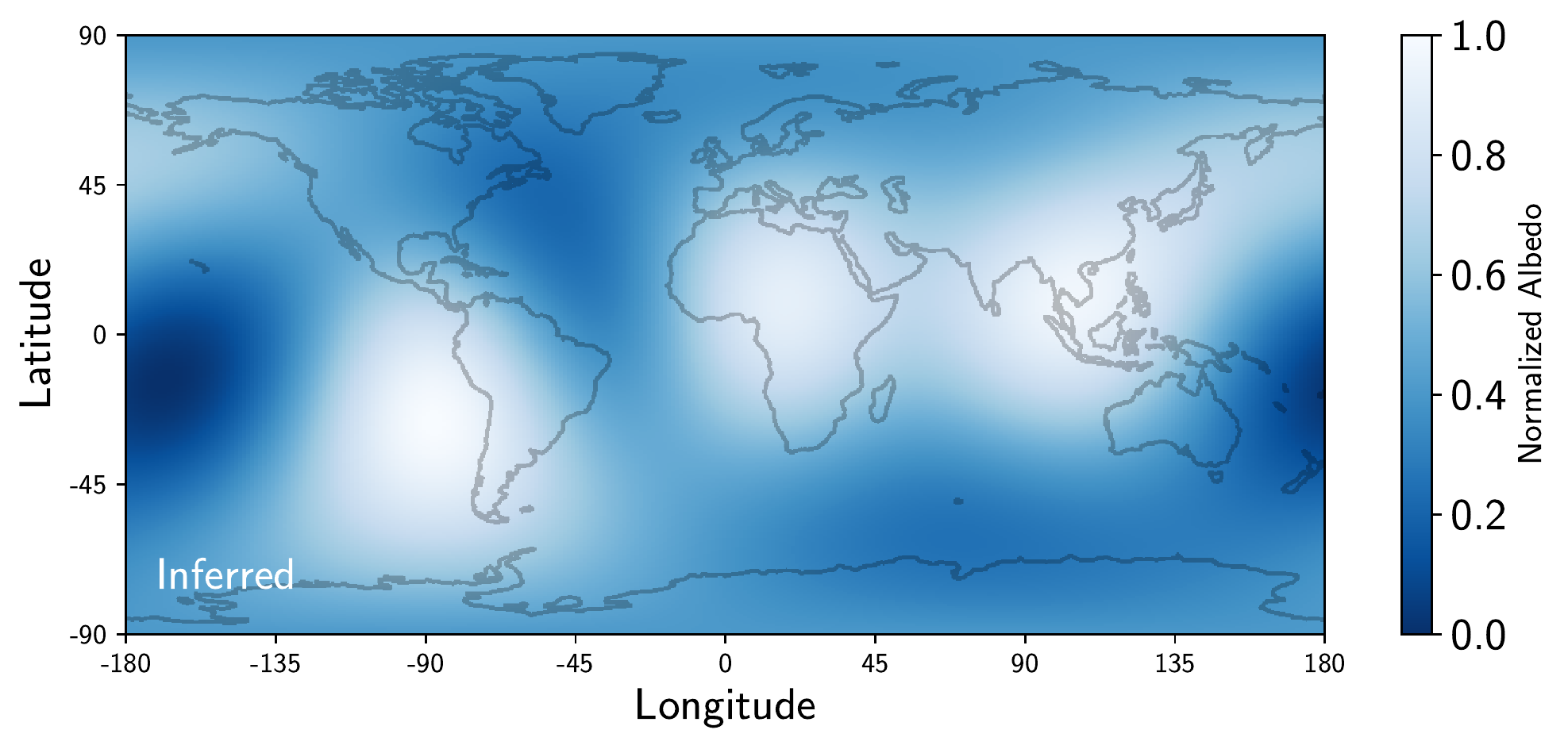}
    \caption{\label{fig:map_L2}
             Maximum likelihood map of Sol d obtained during the first
             stage of our optimization procedure, in which an iterative
             analytic approach was used to solve the bilinear model, 
             subject to very conservative regularization. Despite the low
             resolution of this map, the bright features appear to track
             land masses which for specificity we will refer to as (from left to right)
             ``South America,'' ``Africa,'' and ``Asia.''
             The two dark features are associated with the planet's
             two purported oceans, which we call the ``Pacific'' and the ``Atlantic.'' 
	          Black contours correspond to supposed coastlines
             derived from current state-of-the-art models of the surface geography 
             of Sol d. \codelink{earthshine_S1_S2}      
             }
    \end{centering}
\end{figure}

Figure~\ref{fig:map_L2} shows the maximum likelihood map inferred
during the first stage of our optimization procedure. Since the map
coefficients were heavily regularized, the dynamic range of the albedo
variations was less than 10\%; we scaled it to span the range [0, 1]
for better visualization. Furthermore, although the maximum resolution of
the spherical harmonic model is on 
the order of $18^\circ$, features typically span ${\sim}45^\circ$; this is
also an artefact of our regularization. However, broad features are already
visible in the map: three bright regions and two dark regions (one of which
wraps around the antimeridian). Overplotted on the inferred map are
coastal outlines corresponding to our current best understanding of 
the geography of Sol d. One can see that the three bright regions roughly
track, from left to right, continents we shall refer to as ``South America,''
``Africa,'' and ``Asia,'' in keeping with the literature. The dark regions
are associated with the two major purported oceans on the planet, the
``Pacific'' (centered at longitude $\pm180^\circ$) and ``Atlantic''
(centered at longitude $\-45^\circ$).

\begin{figure}[t!]
    \begin{centering}
    \includegraphics[width=\linewidth]{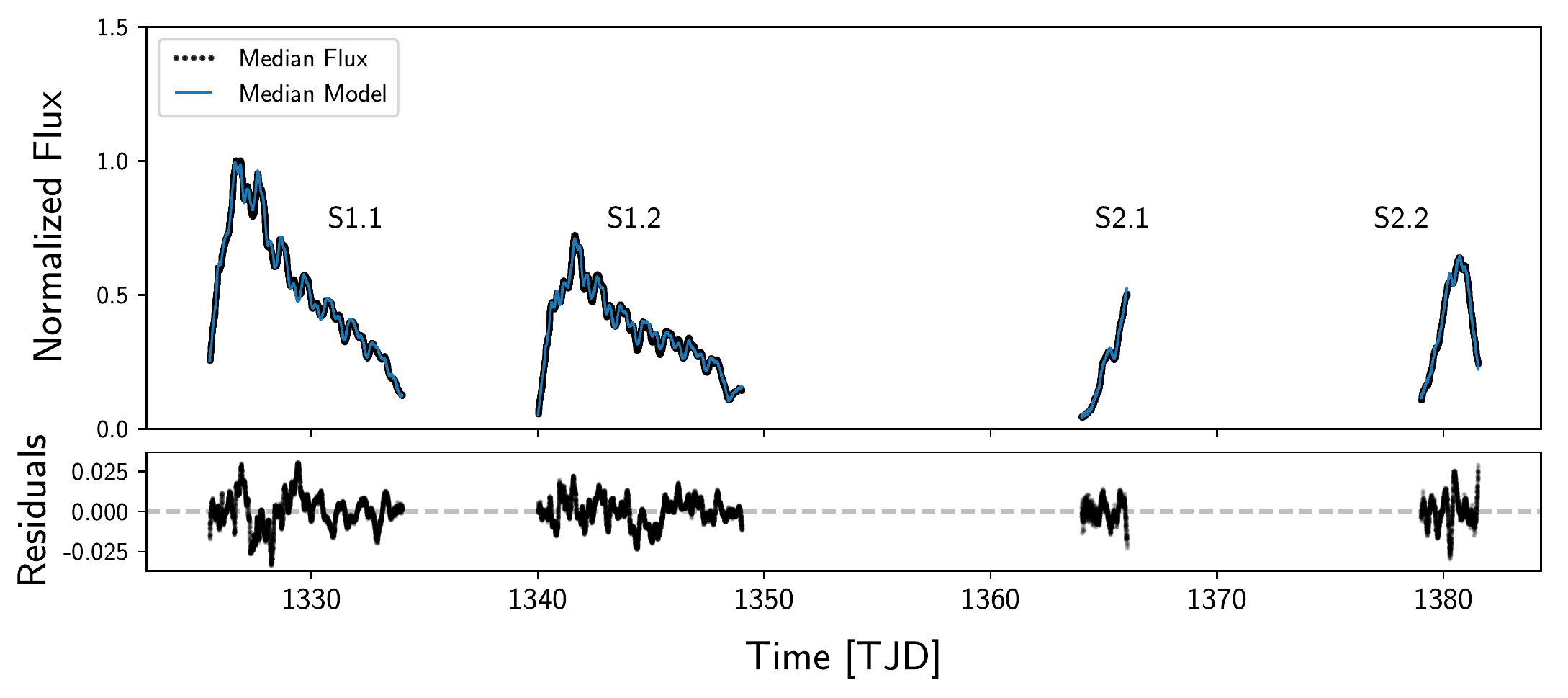}
    \caption{\label{fig:model}
             \emph{Top:} Median flux across all postage stamps (black) and
             the median maximum likelihood model (blue) across Sectors 1 and 2.
             \emph{Bottom:} Median residuals about the maximum likelihood fit.
             \codelink{earthshine_S1_S2}
             }
    \end{centering}
\end{figure}

As expected, the second, non-linear step in our optimization procedure
yields a much better fit to the data.
The top panel of Figure~\ref{fig:model} shows the median of the 
maximum likelihood model in this second step
across all 95 targets (blue) overplotted on the median flux (black). 
The median residuals are shown in the bottom panel; these have
standard deviation of less than one percent. While there is significant correlated
structure in the residuals, their power spectrum displays a broad peak extending 
between about 0.1 and 10 days, with a significant dip at 1.00 day. This
suggests our model is fitting the rotational variability of Sol d quite well, but
additional temporal variations---most likely due to cloud movement---cause aperiodic
changes in the reflectivity of the surface. We revisit this point in \S\ref{sec:discussion}.

\begin{figure}[p!]
    \begin{centering}
    \includegraphics[width=\linewidth]{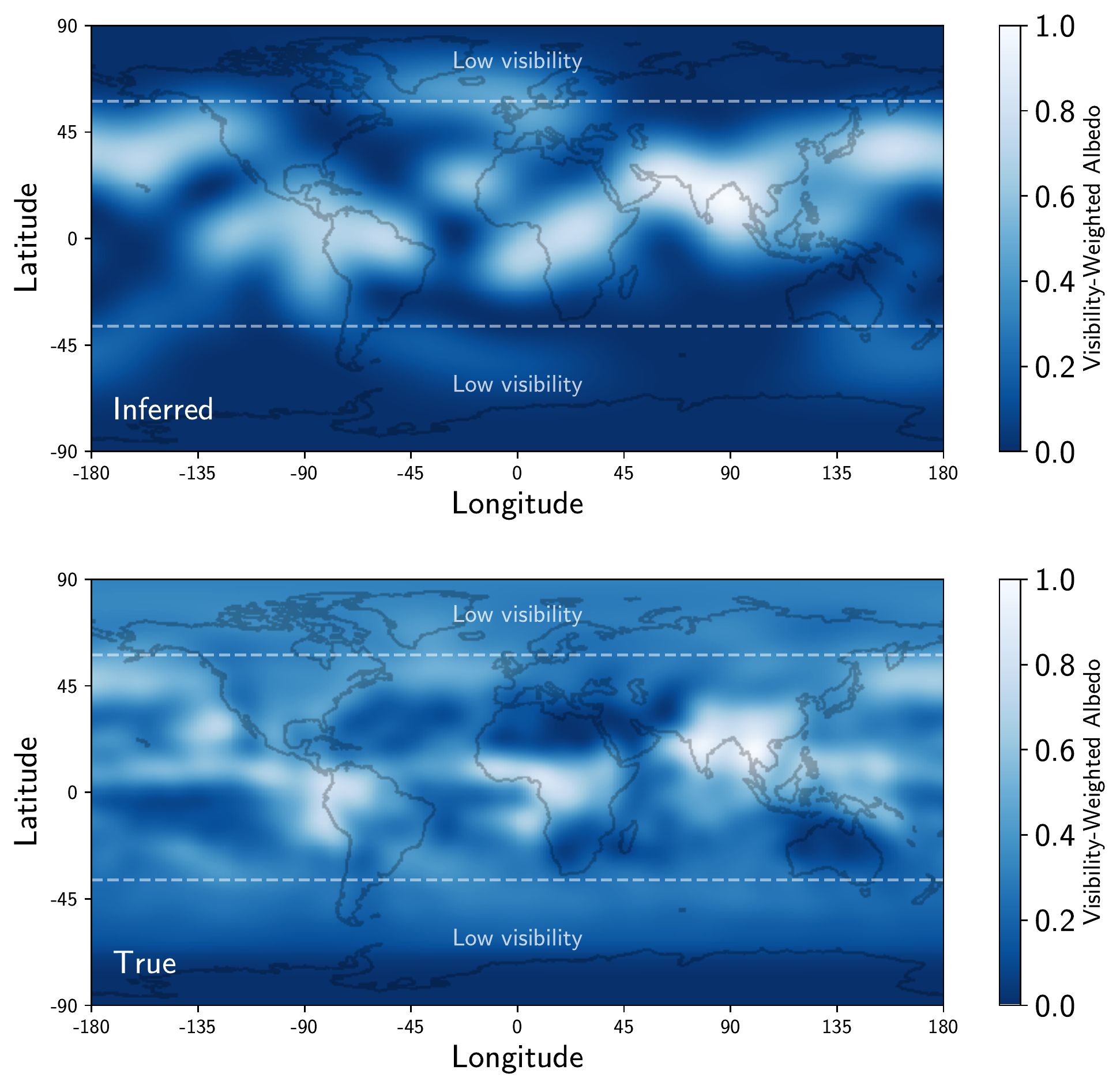}
    \caption{\label{fig:map}
             \emph{Top:} The maximum likelihood model for the surface albedo
             of Sol d, weighted by the visibility across Sectors 1 and 2. The
             dashed white lines indicate latitudes above/below which the
             visibility drops below 50\%. As in Figure~\ref{fig:map_L2}, coastlines 
             based on current models from the literature are overplotted as black contours.
             \emph{Bottom:} Approximate cloud coverage map based on corrected
             surface reflectance obtained by the
             \emph{VIIRS} imager, averaged over the same date range and weighted
             by the same visibility function.
             \codelink{earthshine_S1_S2}
             }
    \end{centering}
\end{figure}

In the top panel of Figure~\ref{fig:map} we show the final,
maximum likelihood albedo map inferred for Sol d weighted by a visibility function.
The visibility function is computed as the average of the product of the 
cosine of the illumination angle and the cosine of the angle between the 
observer vector and the vector normal to the surface of Sol d. This normalization
has the effect of downweighting regions that are either seldom illuminated (such
as regions near the southern pole of Sol d, as the observations were taken during
southern winter) or seldom in view (such as regions near the northern pole of the planet,
which during Sectors 1 and 2 were mostly on the opposite side of Sol d from \TESS).
Because there is little data from these regions, their albedo is prior-dominated
and tends to fluctuate substantially with small changes to the choices we make
when fitting the data. On the other hand, we find that the large scale features close to
the equator of Sol d are mostly insensitive to our choices of prior. In the Figure,
we use dashed lines to delineate the regions whose average visibility is larger
than 0.5.

The bottom panel of Figure~\ref{fig:map} shows an approximate average
cloud cover map of Sol d based on imagery taken by the
Visible Infrared Imaging Radiometer Suite (\emph{VIIRS}) instrument aboard the 
Suomi National Polar-orbiting Partnership (\emph{S-NPP}) satellite%
\footnote{Data obtained via 
\href{https://github.com/rodluger/earthshine/blob/master/tex/figures/viirs.sh}{this wget script}.}. 
The image was produced by analyzing the corrected true color reflectance in the
I1 (red), M4 (blue), and M3 (green) bands and masking
all pixels that were either dark or had significant variance among the three
bands (i.e., inconsistent with a grey spectrum). We then averaged the images
averaged over the 28 days of data analyzed in this work, smoothed it 
with a Gaussian filter, and weighted by the \TESS visibility as in the top panel.

While the two panels display fundamentally different quantities---an albedo
and a cloudiness factor---we expect that the signal of Sol d in the \TESS
bandpass is dominated by cloud reflectivity, as water clouds have an albedo
higher than soil, vegetation, or ocean in the optical/near infrared
\citep[e.g.,][]{Jedlovec2009}. We therefore
expect large, coherent cloud structures to show up as the brightest regions 
in our inferred map, and in fact we find that this is generally the case. The
dominant feature in our inferred map is a bright region spanning longitudes
of $+50^\circ$ to $+100^\circ$ just northward of the equator. This is also the
dominant feature in the VIIRS map, corresponding to the persistent weather system
associated with summertime monsoons in the south of Asia.

Secondary features common to both maps are a permanent cloudy region over
central Africa with a westward spur over the Atlantic Ocean, a veneer of clouds
over the north Pacific, and a pile up of clouds on the western coast of
northern South America. The dark regions in the inferred map rather loosely
track the Pacific, Atlantic, and Indian Oceans, although the correspondence
is not exact.

\section{Discussion}
\label{sec:discussion}

\subsection{An Ill-Posed Problem}
\label{sec:illposed}

The problem of inferring a two-dimensional map from a one-dimensional
time series is famously ill-posed given its large null space and many complex
degeneracies \citep[e.g.,][]{CowanFuentesHaggard2013}. In fact, for a 
uniformly illuminated body, its rotational light curve encodes at most
$2l_\mathrm{max}$ modes (one sine term and one cosine term per frequency), 
where $l_\mathrm{max}$ is the highest spherical
harmonic degree of features on the surface of the body. The number of modes
\emph{on the surface of the body}, however, increases quadratically 
as $(l_\mathrm{max} + 1)^2$. This means that for a body whose surface is
perfectly described by a spherical harmonic map of degree $l_\mathrm{max} = 10$,
which is described by 121 coefficients, only 20 linear combinations
of those coefficients are actually projected onto the light curve. The
problem is fundamentally non-invertible, since most of the information
one wishes to extract never makes it into the light curve in the first place.

Fortunately, there are two aspects of the problem at hand that facilitate
this (seemingly impossible) inversion. First, the surface of Sol d is
not uniformly illuminated; instead, different parts of the surface are weighted
by different amounts as the planet rotates and as the angle of incident starlight
moves over the course of the planet's year. This weighting is sufficient to break many
of the degeneracies, which are rooted in the anti-symmetry of many of the spherical
harmonic modes. Moreover, the discontinuous day/night terminator further
breaks these degeneracies, as only modes that are anti-symmetric about
the visible lune of the planet strictly remain in the null space. Second, 
the changing vantage point of \TESS relative to Sol d (and our exact knowledge of it)
breaks degeneracies that are viewpoint-dependent. In particular, the motion of
\TESS changes the position and shape of the terminator on the projected disk
of Sol d, enhancing the effect discussed above.

In fact, because of these points, there should actually be \emph{no} null space 
in this particular problem. In practice, however, it is evident from
Figure~\ref{fig:map} that degeneracies still abound. While we infer a few of the
dominant features of the cloud map of Sol d, several of them are skewed, 
shifted, or entirely out of place. This is the case, for instance, with
the cloud bank west of South America, which extends too far westward, and
the northern Atlantic Ocean, which appears to extend into the east coast of
North America. We identify four broad reasons for this.

First, many of the issues arise because
of the finite signal-to-noise of our measurements and various biases in
our adopted model. While the measurement errors on our data points are
quite small---since the signal of Sol d is both quite strong and independently
measured at 95 different positions on the detector---our lack of understanding
of the complex optics (and our
attempt to na{\"i}vely model it with pixel-level decorrelation) likely
introduces significant correlated noise when we de-trend. We discuss this
and other sources of bias in more detail in \S\ref{sec:bias} below.

Second, while we attempted to adopt agnostic priors on the map coefficients---i.e.,
we fit for a spectral slope rather than impose one---our inferred map is still 
heavily prior-dependent. Our choice of regularization strength for the systematics
was based on trial-and-error and what ``looked good,'' as was the choice for the 
hyperprior on the spectral
slope and the maximum spherical harmonic degree of the map. Although the main
features of the map (the monsoon over south Asia, the clouds over Africa and
South America, and dark regions associated with the three major oceans) persist
regardless of the choice of prior, their exact shapes and contrast vary
significantly under different assumptions.

Third, inspection of Figure~\ref{fig:phases} shows that most of the time, Sol
d is close to full phase as seen by \TESS.
With the terminator close to the limb
of the planet, its ability to break degeneracies is considerably lessened.
We are therefore somewhat limited in our ability to infer in particular the
latitudinal position of features. Observations of Sol d in upcoming
sectors may contain longer segments in which Sol d is seen in crescent,
which should help mitigate these degeneracies.

Fourth, and most importantly, we have not performed actual inference: that is, 
we simply found the maximum likelihood solution (subject to our constraints),
with no estimate of the uncertainty anywhere on our map. A much better
approach than simply maximizing the likelihood function is to marginalize
over the nuisance parameters---such as the systematic model weights and the spectral
slope. This is more costly, but certainly not intractable, and we will
pursue this in future work.

\subsection{Confirmation Bias}
\label{sec:confirmation}

It is difficult to avoid being disingenuous in one's solution
when the answer (or the expected answer)
to a problem is known ahead of time. Since we had access to 
``ground truth'' from the beginning, many of the de-trending/processing/fitting
choices we made were (however inadvertently) driven by the desire to
come as close as possible to the correct answer. While many of these choices
correspond to numbers we have explicitly quoted here, such as the prior
variances on our model terms, several of them are implicit decisions on
\emph{how} we performed the analysis. For instance, in an early version of 
the project we used as regressors actual \TESS housekeeping variables---such as the angle 
and elevation of the Earth and the Moon relative to each of the four camera 
boresights---instead of the PLD basis. However, the systematics model constructed from these quantities
proved to be too inflexible, judging in part from the poor fidelity of the recovered
maps. The results in this work
must therefore be taken with a grain of salt. That said, we have attempted to
offset the possibility of confirmation bias by making our model
(Equation~\ref{eq:model}) as simple as possible in the end (a multiplicative baseline
and a linear reflectance model) and thus limiting
the number of knobs we could turn. We also opted to forego any use of 
real imaging data in our model; for instance, we fit for the spectral
slope of the map as a free parameter rather than using the true power spectrum 
of Sol d. Moreover, the broad features in the strongly
regularized map (Figure~\ref{fig:map_L2})---and in particular the monsoon clouds
over southeast Asia---are insensitive to most of our model choices.

\subsection{Other Kinds of Bias}
\label{sec:bias}

One of the largest sources of bias is the fact that we assumed our
systematics model is strictly multiplicative. Although we subtracted
an estimate of the baseline non-Sol d background flux from each target
(measured from the mid-sector portions of the light curve where the
planet was below the edge of the sunshade), we implicitly assumed
that this constant baseline was the \emph{only} source of additive
noise. We know this to be strictly untrue, as Sol d is known to
have a companion (Sol d I), a moon about one-third the diameter of 
Sol d, whose reflected light should cause periodic modulations
in the background flux of \TESS pixels. Because of its small size and
very low ($\sim 0.1$) albedo, we chose to ignore its effect, but it
is likely to contaminate our signal at certain points in the orbit
where \TESS is closer to Sol d I than to Sol d. While this will
likely bias the \starry model at those times, the effect is particularly
concerning because of its impact on the PLD basis. While PLD can be an
extremely powerful de-trending technique, it is notorious for
grossly overfitting signals in the presence of uncorrected background
modulations \citep{Luger2016}. This happens because PLD strictly
assumes the noise is multiplicative, which is not the case for 
signals with varying (additive) background levels. In principle, one
could model the reflectance variability of Sol d I in the same way as
we did here, but since the optics are not well understood, this 
may be difficult in practice. 

In our analysis, we also neglected the effects of limb darkening. Because of
atmospheric attenuation, features near the limb of the planet should in
reality contribute less to the flux than our model predicts. This could
bias our results particularly in cases where the planet is seen at half
phase (for instance, during days 1326 and 1341 in Figure~\ref{fig:phases}).
In the absence of limb darkening, the brightest regions on the planet are those
close to the sub-stellar point, which is on the limb. Limb darkening should
shift the regions of peak brightness closer to the center of the planet, adding
an east-west bias in the location of the features we identify.

We also ignored the finite angular size of Sol at Sol d (about $0.5^\circ$),
treating it instead as a point source. Regions immediately nightward of the 
terminator should therefore contribute a small amount of flux, which we neglect.
Similarly, we used the timestamps reported in the \tess light curve
FITS files, which are corrected to the barycenter of the Sol system. Since
our signal originates from a source that is closer to \tess than the
barycenter of the system, the timestamps may be wrong by up to 8 minutes,
the light travel time from \tess to Sol. Over the course of 8 minutes, Sol d rotates
by about $2^\circ$, so there should be a blurring of our inferred map at
this scale. However, these two effects
are well below the resolution of our map, so we expect
them not to introduce significant bias.

Moreover, we implicitly assumed an isotropic prior on the spherical harmonic
coefficients by imposing a power spectrum in only the degree $l$. This has the
effect of lessening the extent to which features can be confined to certain
latitudes. In particular, our prior implicitly disfavors banded cloud structure,
which one may \emph{a priori} expect on a rapidly rotating planet such as 
Sol d. The bottom panel of Figure~\ref{fig:map} shows such banded structure
at the equator and at mid-latitudes due to the combined effects of 
north-south circulation and deflection by the Coriolis force.

Finally, and perhaps more importantly, our model assumed a \emph{static}
map for the surface of Sol d. We also know this to be incorrect, since
clouds form and move on timescales of hours to days. We discuss this in
detail in the following section.

\subsection{Temporal Evolution}
\label{sec:temporal}

As we mentioned in \S\ref{sec:results}, the residuals about our maximum
likelihood fit show correlated structure with power at a large range
of periods ranging from 0.1 up to 10 days. This is likely due to 
the fact that rotational variability is not the sole source of the
signal of Sol d: additionally, there is temporal variation in the 
albedo of certain parts of the map, most probably due to cloud movement.
In order to test this, we used \starry to construct a time-variable
spherical harmonic map, defined by a vector of spherical harmonic coefficients
that varies slowly over time:
\begin{align}
    \mathbf{y}(t) = \sum_{n=0}^N \mathbf{y}_n c_n(t)
\end{align}
where the $\mathbf{y}_n$ are the $N$ spherical harmonic coefficient
vectors that describe the temporal variability and the $c_n$ are the components
of the basis in time. Since this is a linear operation on the spherical
harmonic coefficients, the \starry model is still linear, and may be
solved in the same way as before. 

We choose the $c_n$ vectors to be an orthogonal polynomial basis in time of 
degree 6 spanning all of Sector 1. Under this basis, one can 
interpret
$\mathbf{y}_0$ as the static component of the map, $\mathbf{y}_1$ as the
component that varies (quasi-)linearly in time, $\mathbf{y}_2$ as the component
that varies (quasi-)quadratically, and so forth. In all, we fit for six
of these map components.

We neglect the data taken in Sector 2, focusing instead on the portion of the data for which
we have good temporal coverage. In order to mitigate overfitting due to
the larger number of variables we are solving for,
we add a term to our likelihood function that forces the coefficients
of the $n > 0$ maps to remain small. We further evaluate the albedo of the map
on a grid at six equally spaced points in time, enforcing our albedo and
power spectrum priors separately at each of these points.

\begin{figure}[p!]
    \begin{centering}
    \includegraphics[width=\linewidth]{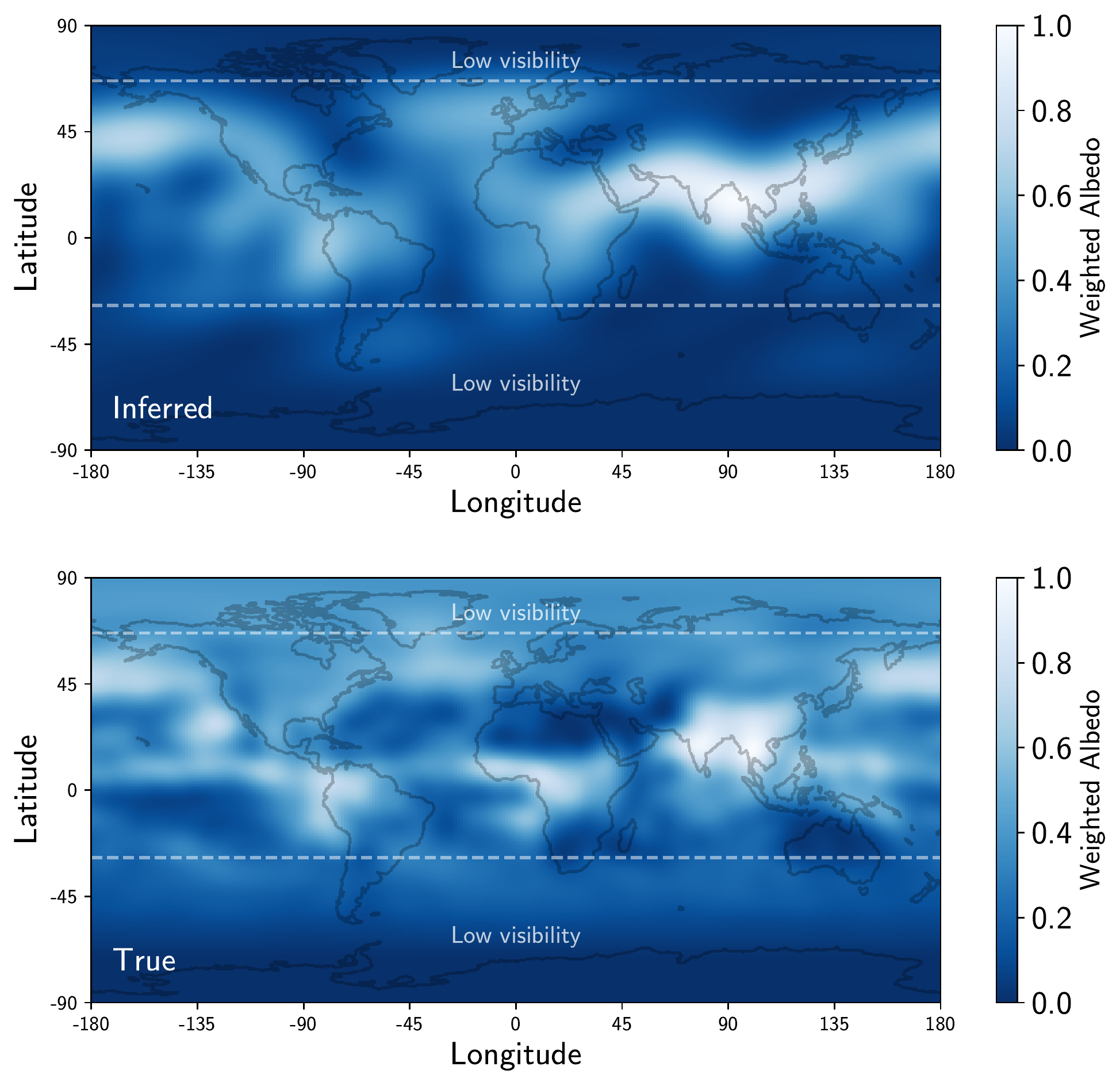}
    \caption{\label{fig:map_temporal}
             The inferred mean map when allowing for slight temporal
             variations in the spherical harmonic coefficients. Compare
             to Figure~\ref{fig:map}. An animated version of this figure
             is available on 
             \href{https://github.com/rodluger/earthshine/blob/master/tex/figures/map_temporal.mp4}{GitHub}.
             \codelink{earthshine_S1_S2}
             }
    \end{centering}
\end{figure}

The mean of the temporal solution is shown in Figure~\ref{fig:map_temporal},
and a link to an animated version is provided in the caption. No significant
differences are evident, although the residuals improved by ${\sim}30\%$
and contain significantly less power on timescales longer than 1 day.
We conclude that while this is evidence for cloud variability on these
timescales, a more detailed investigation is warranted to say anything
meaningful about the dynamics of clouds on Sol d.

\subsection{Implication for Future Observations}
\label{sec:otherplanets}

As we discussed above, light curve inversion is a hard problem,
and we have shown that this is true even at high signal-to-noise and 
when the solution is
known ahead of time. Presented with the top panel of Figure~\ref{fig:map}
(without the continent outlines),
an inhabitant of Sol d may be hard-pressed to identify that
as a map of their home planet. This is in part due to the biases
in our optimization, which we discussed above, but also due to the 
simple fact that the reflectance of Sol d is dominated---by far---by 
time-variable clouds. While stationary cloud features on Sol d are
typically associated with continents (the most pronounced cloud features
being over South America, Africa, and south Asia), persistent cloud
cover can occur over open ocean, as seen in the north Pacific
in Figure~\ref{fig:map}. Given a baseline of just a few weeks, it is
not possible to tell the difference between highly reflective land
features (such as ice or vegetation) and stationary clouds.

We should not expect other exoplanets in the habitable zone to be
any different. Future missions such as LUVOIR or HabEx may enable
us to collect data on the phase curves of such planets, but that
data will be of significantly lower quality and poorer time resolution,
and we will not know the planet's rotation period or obliquity ahead of time.
The work presented here can therefore be seen as an upper limit on what we may
be able to infer about exoplanets in the habitable zone with
next-generation facilities.

That said, the time baseline considered in this work was fairly
limited---three weeks in Sector 1 and one week in Sector 2,
spanning two months in total. Observations taken throughout an entire
year on Sol d---during which time cloud patterns should change with the
seasons and the angle between the day/night terminator and the 
axis of rotation should precess---may help to disambiguate between
clouds, oceans, and continents. Although the signal of Sol d 
mostly disappears from the \tess detector after Sector 2, it is
expected to return in Sectors 14 and 15 when the planet will again
be above the edge of the sunshade.
We intend to perform this analysis
in upcoming work once data from future \TESS sectors becomes available.

Additionally, time-resolved observations in multiple bands could
also help with this problem. This was discussed in \cite{Cowan2009},
who showed that multi-band photometry of the Earth enables one to
disentangle the contributions from oceans, continents, and cloud cover.
A robust detection of surface features on an exoplanet in the habitable
zone will likely necessitate both approaches alongside a circumspect
analysis of the uncertainties and degeneracies at play.

\section{Conclusions}
\label{sec:conclusions}

We have used \TESS photometry to
produce a global map of Sol d, a terrestrial planet in the habitable
zone of a G2 main-sequence star. We adapted the \starry code to 
generate (and invert) light curves in reflected light, coupling it
to a variation of pixel level decorrelation to model the systematics
in the \TESS detector. 

We detect three large scale bright regions near
the equator of the planet, associated with persistent cloud structures
at longitudes of $-60^\circ$, $+30^\circ$, and $+90^\circ$, which
likely track continents. Pronounced dark features are identified
at longitudes of $\pm 180^\circ$, $-30^\circ$, and $+60^\circ$ with
significant latitudinal extent, likely corresponding to oceans. Because of the
vantage point of \TESS during Sectors 1 and 2, we are not sensitive to
high latitudes on the planet, although this may change with upcoming
sectors. 

We fit Sol d with both static and time-variable albedo maps. 
We find tentative evidence for variability in Sol d's cloud coverage over 
the course of Sector 1 observations. 
Upcoming observations with \TESS will yield further information on 
the time variability of surface features on Sol d on a variety of timescales.

\begin{figure}[t!]
    \begin{centering}
    \includegraphics[width=0.4\linewidth]{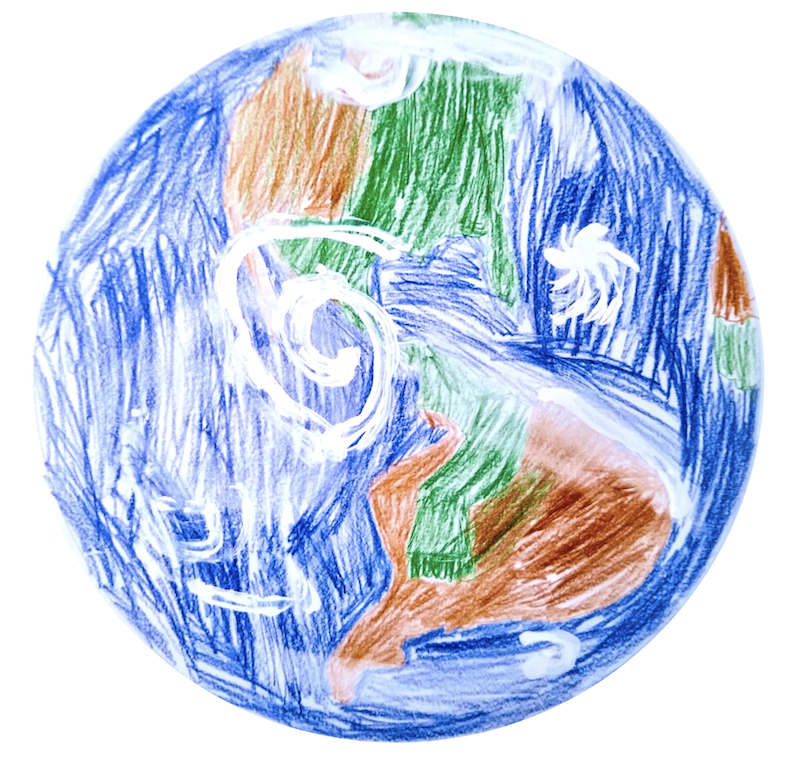}
    \caption{\label{fig:artist}
             Artist's conception of Sol d.
             \emph{Credit: Ian Peng-Sue, age 7.}
             }
    \end{centering}
\end{figure}

To summarize, we have robustly detected surface features from \TESS 
light curves of Sol d. While the resolution of the map we obtain is limited, 
it is clear that surface features including low-albedo oceanic regions and 
high-albedo cloud banks or land masses are present on this terrestrial 
world. An artist's impression of the planet is shown in Figure~\ref{fig:artist}. 
While somewhat fanciful and rather far-fetched, this artistic interpretation 
is nevertheless broadly consistent with the data.

The likely prevalence of continents, oceans, and complex weather patterns 
on Sol d make it an extremely promising site for the evolution of life. 
We strongly advocate for future missions to further investigate this possibility. 
In particular, Sol d may be a good choice of target for a lander mission. 
One prime landing site is located at the planetary (latitude, longitude) coordinates 
(\href{https://www.google.com/maps/place/Roswell,+NM/@36.2909732,-94.0798131,3.96z/data=!4m5!3m4!1s0x86e2651365aced55:0xe41b0be474cfd77e!8m2!3d33.3942655!4d-104.5230242}{$33.3943^\circ$ N, $104.5230^\circ$ W}).%
\footnote{We considered the alternative landing site of southern Oceania, but rumors of murderous native fauna including the fearsome ``drop bear'' (Andrew R. Casey, personal communication) led us to conclude that this would not be a suitable location.} 

The work presented here also serves as a relevant case study for future 
efforts to characterize terrestrial exoplanets. While the circumstances of \TESS's 
observations of Sol d will not be replicated exactly for other planets, similar 
rotationally-modulated phase curves of reflected light may be feasible in the future. 
We show that a fast, analytic linear model can be successfully applied to such data. 
We also underscore the importance of high cadence \emph{and} long duration for 
these observations, and we advocate for multi-wavelength coverage to 
help disentangle the photometric signatures of continents and persistent cloud features.

In conclusion, we have demonstrated that \TESS is an effective addition to NASA's fleet of weather satellites.

\vspace{0.5in}

\acknowledgements{%
Our heartfelt thanks go to Ian Peng-Sue and his agent Stephanie Tonnesen for 
Figure~\ref{fig:artist}. We also gratefully acknowledge Jonathan Fraine, Dan Foreman-Mackey, 
David W. Hogg, and Ben Pope for useful conversations. 
Crucial parts of this work were carried out at the 
TESS Data Workshop, hosted by Space Telescope Science Institute, and the 
Building Early Science with TESS Workshop, hosted by the University of Chicago.}
\facility{TESS}
\software{Astropy, Matplotlib, Numpy, \starry \citep{Luger2019}, 
TensorFlow \citep{Abadi2015}, \spiceypy \citep{Annex2019}}

\newpage
\bibliography{bib}

\end{document}